\documentclass[twocolumn,showpacs,preprintnumbers,amsmath,amssymb]{revtex4}

\usepackage{graphics}      
\usepackage{graphicx}      
\usepackage{url}           
\usepackage{bm}            
\usepackage{amsmath}
\usepackage{xcolor}

\usepackage{tikz}
\usetikzlibrary{decorations.pathmorphing,patterns}

\usepackage{fancyhdr}
\begin{document}
\title {Quasi-stable Localized Excitations in the $\beta$-Fermi Pasta Ulam Tsingou System}
\author {Nathaniel J. Fuller}
\email[Corresponding Author: ]{nfuller2@buffalo.edu}
\author {Surajit Sen}
\email[]{sen@buffalo.edu}
\affiliation {Department of Physics, University at Buffalo - SUNY, Buffalo, New York 14260-1500}
\date{\today}

\begin{abstract}
The lifetimes of localized nonlinear modes in both the $\beta$-Fermi-Pasta-Ulam-Tsingou ($\beta$-FPUT) chain and a cubic $\beta$-FPUT lattice are studied as functions of perturbation amplitude, and by extension, the relative strength of the nonlinear interactions compared to the linear part. We first recover the well known result that localized nonlinear excitations (LNEs) produced by a bond squeeze can be reduced to an approximate two-frequency solution and then show that the nonlinear term in the potential can lead to the production of secondary frequencies within the phonon band. This can affect the stability and lifetime of the LNE by facilitating interactions between the LNE and a low energy acoustic background which can be regarded as ``noise'' in the system. In the one dimensional FPUT chain, the LNE is stabilized by low energy acoustic emissions at early times; in some cases allowing for lifetimes several orders of magnitude larger than the oscillation period. The longest lived LNEs are found to satisfy the parameter dependence $\mathcal{A}\sqrt{\beta}\approx1.1$ where $\beta$ is the relative nonlinear strength and $\mathcal{A}$ is the displacement amplitude of the center particles in the LNE. In the cubic FPUT lattice, the LNE lifetime $T$ decreases rapidly with increasing amplitude $\mathcal{A}$ and is well described by the double log relationship $\log_{10}\log_{10}(T)\approx -(0.15\pm0.01)\mathcal{A}\sqrt{\beta}+(0.62\pm0.02)$.
\end{abstract}


\maketitle
\section{Introduction}
The formation of highly localized bundles of energy seems to be a universal feature of discrete lattices with nonlinear couplings \cite{Jackson1990}. These energy packets, termed LNEs, likely play a fundamental role in the transmission of energy through anharmonic crystals and polymers as well as the long-term relaxation of the system to thermal equilibrium. LNEs have been observed experimentally in micromechanical oscillator arrays with lifetimes that are several thousand times larger than their oscillation periods \cite{Sato2003}. LNEs appear across a broad range of systems where both nonlinearity and discreteness contribute to the system dynamics, some examples include micromechanical oscillator arrays \cite{Sato2003}, antiferromagnetic spin lattices \cite{Sato2004}, organic conducting polymers \cite{Tretiak2004}, halide-bridged transition metal compounds \cite{Swanson1999}, high-temperature lattice excitations in NaI \cite{Manley2009}, protein conformational dynamics \cite{Xie2000,Xie2001,Luccioli2011,Nicolai2015}, site specific enzymatic activity \cite{Juanico2007,Piazza2009}, and Bose-Einstein condensates \cite{Trombettoni2001}.
\par
Unfortunately, a complete understanding of how LNEs behave throughout the dynamical evolution of a nonlinear lattice is yet to be developed \cite{Kashyap2017}. Prior studies have generally focused on finding the envelope function of a breather solution to probe the dynamics of localized excitations \cite{Brickham1993, Kosevich1993_2, Flach1993, Flach1994, Fuller2020}. These breather solutions are a special class of LNEs where the motion of the system is exactly periodic, meaning that the set of frequencies present within the breather are commensurate. This allows the LNE solution to be described in terms of trigonometric or elliptic functions where the resulting behavior is closely related to the phenomena of localized nonlinear modes. However, there does not appear to be a fundamental requirement for LNEs in general to be exactly periodic. This lack of periodicity can complicate the stability analysis of LNEs since the standard approaches such as Floquet theory test the stability of perturbed periodic solutions. Furthermore, it is likely that the absence of exactly periodic motion is closely related to the eventual break down of LNEs and approach to thermal equilibrium observed in simulations on finite lattices.
\par
Considering these points, we will forgo searching for exact breather solutions here and instead focus our study on LNEs created by simple initial conditions, such as a bond squeeze, which can produce long lived excitations that are not exactly periodic and eventually break down as the system approaches equilibrium. It should be stressed however, that instability does not imply that the LNEs are short-lived. An interesting feature of LNEs is their ability to maintain energy localization on time scales many orders of magnitude longer than the oscillation period \cite{Sato2003,Mohan2011}.
\par
Our goal is to both describe the long-term dynamics of the LNEs observed as functions of the system parameters and provide insight into the underlying physical mechanisms which appear to control the lifetime of the LNEs and hence how quickly the system approaches thermal equilibrium. We will use the well known $\beta$-FPUT chain for our model system since it defines both harmonic and nonlinear potential terms in a simple manner. Specifically, the model consists of a one dimensional harmonic oscillator chain with a symmetric fourth order potential added on which can be understood as the first two terms of a power series expansion of a general symmetric nonlinear potential. The long-term dynamics of the LNEs are explored by numerical simulations. We use both the NVE integrator in LAMMPS \cite{Plimpton1995} and the 5th-order Gear integrator in PULSEDYN \cite{Kashyap2019} to run our simulations. This redundancy can help minimize the possibility that a particular solver is giving incorrect results. We find that both methods give similar results for the LNE lifetimes, amplitudes, and frequencies studied here.

\section{The Model}
By appropriately scaling time, the equations of motion for the system can be written in the following simplified form,
\begin{equation}
\ddot{x}_{n}=x_{n+1}-2x_{n}+x_{n-1}+\beta\left[\left(x_{n+1}-x_{n}\right)^{3}+\left(x_{n-1}-x_{n}\right)^{3}\right],
\label{fput}
\end{equation}
where the parameter $\beta$ sets the relative strength of the nonlinearity. Throughout this work, we will be concerned with the dynamics of LNEs where the energy is concentrated among only a few particles; having mode profiles with alternating sign. It is therefore beneficial to introduce the parameter $r_{n}=(-1)^{n}\sqrt{\beta}\left(x_{n+1}-x_{n}\right)$ which combined with Eq. (\ref{fput}) produces,
\begin{equation}
\ddot{r}_{n}+r_{n+1}+2r_{n}+r_{n-1}+r_{n+1}^{3}+2r_{n}^{3}+r_{n-1}^{3}=0.
\label{eom}
\end{equation}
\par
LNEs can be generated by either waiting for them to arise naturally during the dynamical evolution of a non-equilibrium state \cite{Cretegny1998,Westley2018} or by initiating them from specific initial conditions \cite{Kashyap2017,Mohan2011,Fuller2020}. We choose the latter, following the referenced works, by either displacing a set of chosen particles from equilibrium or by giving them an initial velocity perturbation, while in both cases, the rest of the chain is unperturbed. The resulting LNE will be \textit{even-parity} if $x_{1}(0)=-x_{0}(0)=A/(2\sqrt{\beta})$ with $r_{0}(0)=2r_{1}(0)=2r_{-1}(0)=A$ or $\dot{x}_{1}(0)=-\dot{x}_{0}(0)=V/(2\sqrt{\beta})$ with $\dot{r}_{0}(0)=2\dot{r}_{1}(0)=2\dot{r}_{-1}(0)=V$, and likewise will be \textit{odd-parity} if $x_{0}(0)=A/\sqrt{\beta}$ with $r_{-1}(0)=r_{0}(0)=A$ or $\dot{x}_{0}(0)=V/\sqrt{\beta}$ with $\dot{r}_{-1}(0)=\dot{r}_{0}(0)=V$. Here, we have set $n=0$ corresponding to the position of the LNE where $A$ and $V$ are constants related to the initial amplitude and velocity respectively.
\par
We find via dynamical simulations \cite{Kashyap2019} that the LNEs excited using the initial conditions described above tend to consist of two major frequencies stemming from the effective two degrees of freedom afforded to the LNE \cite{Flach1993, Flach1994_2} along with a collection of higher harmonics. The long lived nature of the LNE implies that the system is behaving approximately as a reduced, two degree of freedom system, at least until the onset of delocalization. Of course, this simplification can only be taken so far. The LNE does radiate some energy into the rest of the chain, and it is this radiation, even if it is weak, that ultimately causes the LNE to break down.

\begin{figure}[b]
	\centering
	\begin{tikzpicture}[scale=0.9]
	\coordinate (a1) at (-3.5,0);
	\node[circle,fill=gray,inner sep=2.5mm] (a) at (-3,0) {};
	\coordinate (a2) at (-2.5,0);
	\coordinate (b1) at (-1.5,0);
	\node[circle,fill=gray,inner sep=2.5mm] (b) at (-1,0) {};
	\coordinate (b2) at (-0.5,0);
	\coordinate (c1) at (0.5,0);
	\node[circle,fill=gray,inner sep=2.5mm] (c) at (1,0) {};
	\coordinate (c2) at (1.5,0);
	\coordinate (d1) at (2.5,0);
	\node[circle,fill=gray,inner sep=2.5mm] (d) at (3,0) {};
	\coordinate (d2) at (3.5,0);
	\draw (a1) -- (a);
	\draw (a) -- (a2);
	\draw (b1) -- (b);
	\draw (b) -- (b2);
	\draw (c1) -- (c);
	\draw (c) -- (c2);
	\draw (d1) -- (d);
	\draw (d) -- (d2);
	\draw[decoration={aspect=0.3, segment length=1mm, amplitude=2mm,coil},decorate] (-4,0) -- (a1);
	\draw[decoration={aspect=0.3, segment length=1mm, amplitude=2mm,coil},decorate] (a2) -- (b1);
	\draw[decoration={aspect=0.3, segment length=1mm, amplitude=2mm,coil},decorate] (b2) -- (c1);
	\draw[decoration={aspect=0.3, segment length=1mm, amplitude=2mm,coil},decorate] (c2) -- (d1);
	\draw[decoration={aspect=0.3, segment length=1mm, amplitude=2mm,coil},decorate] (d2) -- (4,0);
	\draw[->,line width=0.5mm,>=stealth] (-1.3,0.6) -- (-0.7,0.6);
	\draw[->,line width=0.5mm,>=stealth] (1.3,0.6) -- (0.7,0.6);
	\node[below] at (0,1.5) {Even Parity LNE};
	\node[below] at (-3,-0.5) {$n=-1$};
	\node[below] at (-1,-0.5) {$n=0$};
	\node[below] at (1,-0.5) {$n=1$};
	\node[below] at (3,-0.5) {$n=2$};
	\node[above] at (-3,0.5) {$x_{-1}\approx0$};
	\node[above] at (3,0.5) {$x_{2}\approx0$};
	
	\node[circle,fill=gray,inner sep=2.5mm] (a) at (-4,-3.5) {};
	\coordinate (a2) at (-3.5,-3.5);
	\coordinate (b1) at (-2.5,-3.5);
	\node[circle,fill=gray,inner sep=2.5mm] (b) at (-2,-3.5) {};
	\coordinate (b2) at (-1.5,-3.5);
	\coordinate (c1) at (-0.5,-3.5);
	\node[circle,fill=gray,inner sep=2.5mm] (c) at (0,-3.5) {};
	\coordinate (c2) at (0.5,-3.5);
	\coordinate (d1) at (1.5,-3.5);
	\node[circle,fill=gray,inner sep=2.5mm] (d) at (2,-3.5) {};
	\coordinate (d2) at (2.5,-3.5);
	\coordinate (e1) at (3.5,-3.5);
	\node[circle,fill=gray,inner sep=2.5mm] (e) at (4,-3.5) {};
	\draw (a) -- (a2);
	\draw (b1) -- (b);
	\draw (b) -- (b2);
	\draw (c1) -- (c);
	\draw (c) -- (c2);
	\draw (d1) -- (d);
	\draw (d) -- (d2);
	\draw (e1) -- (e);
	\draw[decoration={aspect=0.3, segment length=1mm, amplitude=2mm,coil},decorate] (a2) -- (b1);
	\draw[decoration={aspect=0.3, segment length=1mm, amplitude=2mm,coil},decorate] (b2) -- (c1);
	\draw[decoration={aspect=0.3, segment length=1mm, amplitude=2mm,coil},decorate] (c2) -- (d1);
	\draw[decoration={aspect=0.3, segment length=1mm, amplitude=2mm,coil},decorate] (d2) -- (e1);
	\draw[->,line width=0.5mm,>=stealth] (-1.7,-2.9) -- (-2.3,-2.9);
	\draw[->,line width=0.5mm,>=stealth] (-0.3,-2.9) -- (0.3,-2.9);
	\draw[->,line width=0.5mm,>=stealth] (2.3,-2.9) -- (1.7,-2.9);
	\node[below] at (0,-2) {Odd Parity LNE};
	\node[below] at (-4,-4) {$n=-2$};
	\node[below] at (-2,-4) {$n=-1$};
	\node[below] at (0,-4) {$n=0$};
	\node[below] at (2,-4) {$n=1$};
	\node[below] at (4,-4) {$n=2$};
	\node[above] at (-4,-3) {$x_{-2}\approx0$};
	\node[above] at (4,-3) {$x_{2}\approx0$};
	\end{tikzpicture}
	\caption{Approximate motion of an even and odd parity LNE. The arrows above the particles show the direction of the instantaneous motion. Particles with no arrows are considered outside the LNE and approximated as being at rest.}
	\label{parity}
\end{figure}
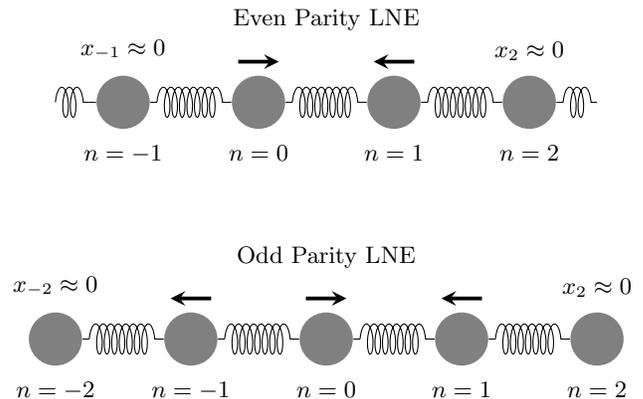

\par
Motivated by the numerical results, we consider the motion of the LNE to consist of two dominant frequencies $\omega_{a}$ and $\omega_{b}$ and that the LNE is confined to $r_{0}$ and its immediate neighbors. The even and odd parity motion of the LNE considered under these assumptions is illustrated in Fig. \ref{parity}. The motion of the even-parity LNE can then be written as,
\begin{equation}
\begin{bmatrix}
r_{-1} \\
r_{0} \\
r_{1}
\end{bmatrix}
=
\begin{bmatrix}
a_{1} & b_{1} \\
a_{0} & b_{0} \\
a_{1} & b_{1}
\end{bmatrix}
\begin{bmatrix}
\cos(\omega_{a}t) \\
\cos(\omega_{b}t)
\end{bmatrix},
\label{ep_ansatz}
\end{equation}
while for the the odd-parity LNE,
\begin{equation}
\begin{bmatrix}
r_{-2} \\
r_{-1} \\
r_{0} \\
r_{1}
\end{bmatrix}
=
\begin{bmatrix}
a_{1} & b_{1} \\
a_{0} & b_{0} \\
a_{0} & b_{0} \\
a_{1} & b_{1}
\end{bmatrix}
\begin{bmatrix}
\cos(\omega_{a}t) \\
\cos(\omega_{b}t)
\end{bmatrix}.
\label{op_ansatz}
\end{equation}
We have ignored possible phase shifts in Eqs. (\ref{ep_ansatz} \& \ref{op_ansatz}) since they appear to be quite small in our simulations and add complexity to the solution which does not lead to fundamentally different results. We generally consider $\omega_{a}$ as the stronger and higher frequency in the center bond at $r_{0}$ with $\omega_{b}$ as the weaker and lower frequency. In both cases, the symmetry of the system reduces the number of coefficients to four. The primary difference is that the even-parity LNEs are (particle) bond centered in ($r$) $x$ while odd-parity LNEs are (bond) particle centered in ($r$) $x$.
\par
Inserting Eqs. (\ref{ep_ansatz} \& \ref{op_ansatz}) into Eq. (\ref{eom}) gives,
\begin{widetext}
\begin{align}
\begin{split}
&\left[2a_{1}+\left(2-\omega_{a}^{2}\right)a_{0}\right]\cos(\omega_{a}t)+\left[2b_{1}+\left(2-\omega_{b}^{2}\right)b_{0}\right]\cos(\omega_{b}t)+\\
&2\left[a_{1}\cos(\omega_{a}t)+b_{1}\cos(\omega_{b}t)\right]^{3}+2\left[a_{0}\cos(\omega_{a}t)+b_{0}\cos(\omega_{b}t)\right]^{3}=0,
\end{split}
\end{align}
\begin{align}
\begin{split}
&\left[a_{1}+\left(3-\omega_{a}^{2}\right)a_{0}\right]\cos(\omega_{a}t)+\left[b_{1}+\left(3-\omega_{b}^{2}\right)b_{0}\right]\cos(\omega_{b}t)+\\
&\left[a_{1}\cos(\omega_{a}t)+b_{1}\cos(\omega_{b}t)\right]^{3}+3\left[a_{0}\cos(\omega_{a}t)+b_{0}\cos(\omega_{b}t)\right]^{3}=0.
\end{split}
\end{align}
The cubic terms can be expanded into their different frequency components using the relation,
\begin{align}
\begin{split}
[a_{n}\cos(\omega_{a}t)+b_{n}\cos(\omega_{b}t)]^{3}=&\frac{1}{4}a_{n}^{3}\cos(3\omega_{a}t)+\frac{1}{4}b_{n}^{3}\cos(3\omega_{b}t)+\left(\frac{3}{4}a_{n}^{3}+\frac{3}{2}a_{n}b_{n}^{2}\right)\cos(\omega_{a}t)+\left(\frac{3}{4}b_{n}^{3}+\frac{3}{2}a_{n}^{2}b_{n}\right)\cos(\omega_{b}t)+\\
&\frac{3}{4}a_{n}b_{n}^{2}\left[\cos([\omega_{a}-2\omega_{b}]t)+\cos([\omega_{a}+2\omega_{b}]t)\right]+\frac{3}{4}a_{n}^{2}b_{n}\left[\cos([\omega_{b}-2\omega_{a}]t)+\cos([\omega_{b}+2\omega_{a}]t)\right].
\end{split}
\label{cubic_exp}
\end{align}
\end{widetext}
For now, we will restrict our attention to the main frequencies $\omega_{a}$ and $\omega_{b}$ by appealing to the rotating wave approximation (RWA) where the contribution to the dynamics from frequencies beyond a predetermined threshold are considered negligible \cite{Kosevich1993_2,Sievers1988,Khomeriki2002}. Considering that our initial guess for the dynamics only contains two frequencies, it is reasonable to disregard higher multiples of these frequencies since doing otherwise would imply that the proposed ansatz is insufficient. To insure harmonic balance is maintained, the coefficients for the even-parity LNE must satisfy,
\begin{subequations}
\begin{align}
2a_{1}+(2-\omega_{a}^{2})a_{0}+\frac{3}{2}\left(a_{1}^{3}+a_{0}^{3}+2a_{1}b_{1}^{2}+2a_{0}b_{0}^{2}\right)=0,\\
2b_{1}+(2-\omega_{b}^{2})b_{0}+\frac{3}{2}\left(b_{1}^{3}+b_{0}^{3}+2a_{1}^{2}b_{1}+2a_{0}^{2}b_{0}\right)=0,
\end{align}
\label{ep_coef}
\end{subequations}
while for the odd-parity LNE,
\begin{subequations}
\begin{align}
a_{1}+(3-\omega_{a}^{2})a_{0}+\frac{3}{4}\left(a_{1}^{3}+3a_{0}^{3}+2a_{1}b_{1}^{2}+6a_{0}b_{0}^{2}\right)=0,\\
b_{1}+(3-\omega_{b}^{2})b_{0}+\frac{3}{4}\left(b_{1}^{3}+3b_{0}^{3}+2a_{1}^{2}b_{1}+6a_{0}^{2}b_{0}\right)=0.
\end{align}
\label{op_coef}
\end{subequations}
\par
For notational convenience, we introduce the variables $\Lambda_{a+}=\frac{2}{3}(2-\omega_{a}^{2})a_{0}+a_{0}^{3}+2a_{0}b_{0}^{2}$ and $\Lambda_{b+}=\frac{2}{3}(2-\omega_{b}^{2})b_{0}+b_{0}^{3}+2a_{0}^{2}b_{0}$ for the even parity LNEs and $\Lambda_{a-}=\frac{4}{3}(3-\omega_{a}^{2})a_{0}+3\left(a_{0}^{3}+2a_{0}b_{0}^{2}\right)$ and $\Lambda_{b-}=\frac{4}{3}(3-\omega_{b}^{2})b_{0}+3\left(b_{0}^{3}+2a_{0}^{2}b_{0}\right)$ for the odd-parity LNEs, which depend only on the parameters in $r_{0}$. This reduces the set of Eqs. (\ref{ep_coef} \& \ref{op_coef}) to the following pair of equations which are valid for both types of LNEs:
\begin{subequations}
\begin{align}
a_{1}^{3}+2\left(\frac{2}{3}+b_{1}^{2}\right)a_{1}+\Lambda_{a\pm}=0,\label{cub_amp_eqs1}\\
b_{1}^{3}+2\left(\frac{2}{3}+a_{1}^{2}\right)b_{1}+\Lambda_{b\pm}=0.\label{cub_amp_eqs2}
\end{align}
\end{subequations}
This yields the amplitudes at the neighboring sites,
\begin{subequations}
\begin{align}
a_{1}=\sqrt{-\frac{\Lambda_{b\pm}+b_{1}^{3}}{2b_{1}}-\frac{2}{3}},
\label{a_rwa}\\
b_{1}=\sqrt{-\frac{\Lambda_{a\pm}+a_{1}^{3}}{2a_{1}}-\frac{2}{3}}.
\label{b_rwa}
\end{align}
\end{subequations}
Note that only the positive solutions are taken in Eqs. (\ref{a_rwa} \&\ref{b_rwa}) since $a_{n}$ \& $b_{n}$ are considered positive numbers; this removes two solutions from the set of cubic equations (\ref{cub_amp_eqs1} \& \ref{cub_amp_eqs2}).
\par
At this point we should note that since $a_{1}$ and $b_{1}$ must be real, Eqs. (\ref{a_rwa} \& \ref{b_rwa}) place a constraint on the allowed frequencies and amplitudes of the LNE. This can be combined with another constraint, where the frequencies $\omega_{a}$ and $\omega_{b}$ must be above the phonon band, to give the minimum LNE amplitude for maintaining energy localization. The phonon band for Eq. (\ref{fput}) spans the frequencies from $0$ to $2$ and is fixed since the linear strength is scaled to one. The upper frequency limit appears due to the discreteness of the system where higher frequencies would correspond to modes with wave numbers smaller than the particle spacing. We see from numerical simulations \cite{Kashyap2019} that as the amplitude of the LNE is lowered, the lower frequency $\omega_{b}$ skims the top of the phonon band; eventually becoming indistinguishable from $\omega_{a}$ as the upper frequency likewise approaches the phonon band. This leads to the inequality,
\begin{equation}
\Lambda_{a\pm}\leq -a_{1}^{3}-\frac{4}{3}a_{1}.
\label{real_req}
\end{equation}
For the even-parity LNE, inserting the initial condition $r_{0}(0)=2r_{1}(0)=A$ into Eqs. (\ref{ep_ansatz} \& \ref{real_req}) gives a minimum amplitude of $a_{min(+)}=\frac{4}{3\sqrt{3}}$, Repeating this for the odd-partiy LNE gives a minimum amplitude of $a_{min(-)}=\frac{2}{3}$. In the context of Eq. (\ref{fput}), the minimum energy of the even-parity (odd-parity) LNE is $\frac{44}{81}$ ($\frac{52}{81}$). This shows that even-parity LNEs generally require less energy to excite than odd-parity LNEs, and helps to justify their improved stability over their odd-parity counterparts \cite{Cretegny1998,Mohan2011, Kivshar1993}.
\par
\begin{figure}[t]
\centering
\includegraphics[width=\columnwidth]{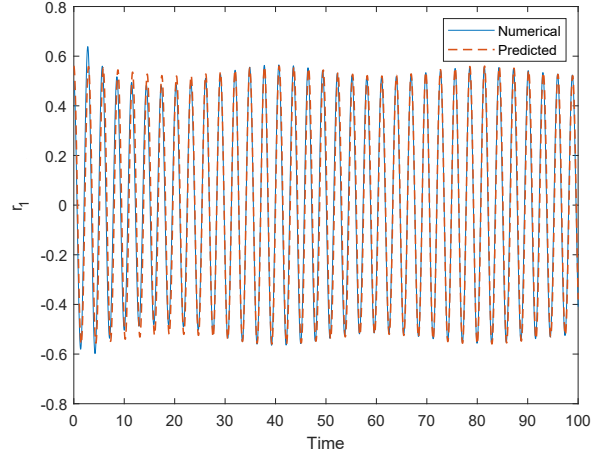}
\caption{The motion of the LNE at $r_{1}$ is shown from a numerically generated solution and the predicted motion from Eqs. (\ref{a1_eq} \& \ref{b1_eq}). The agreement improves after an initial transient period lasting until about $t\approx20$.}
\label{r1_fit}
\end{figure}

Returning to Eqs. (\ref{a_rwa} \& \ref{b_rwa}), the amplitudes can be decoupled to produce the following equations for $a_{1}$ \& $b_{1}$:
\begin{align}
\left(\frac{3}{4}a_{1}^{3}+\frac{1}{3}a_{1}-\frac{\Lambda_{a\pm}}{4}\right)^{2}\left(2a_{1}^{3}+\frac{8}{3}a_{1}+2\Lambda_{a\pm}\right)+\Lambda_{b\pm}^{2}a_{1}^{3}&=0,
\label{a1_eq}\\
\left(\frac{3}{4}b_{1}^{3}+\frac{1}{3}b_{1}-\frac{\Lambda_{b\pm}}{4}\right)^{2}\left(2b_{1}^{3}+\frac{8}{3}b_{1}+2\Lambda_{b\pm}\right)+\Lambda_{a\pm}^{2}b_{1}^{3}&=0.
\label{b1_eq}
\end{align}
Eqs. (\ref{a1_eq} \& \ref{b1_eq}) can be solved numerically for a known $r_{0}$ to give the motion of the oscillator adjacent to the LNE at $r_{1}$. For example, we consider an even-parity LNE initiated by the perturbation $r_{0}(0)=1$ where the scaling for $r$ suggests that the linear and nonlinear forces are commensurate at this amplitude. The LNE shows motion in the central bond corresponding to Eq. (\ref{ep_ansatz}) with $a_{0}\approx0.64$, $b_{0}\approx0.058$, $\omega_{a}\approx2.160$, and $\omega_{b}\approx2.005$. The solution to Eqs. (\ref{a1_eq} \& \ref{b1_eq}) with those parameters is $a_{1}\approx0.54$ and $b_{1}\approx0.02$. A comparison between the numerical solution and predicted motion of the neighboring oscillator is shown in Fig. \ref{r1_fit}. We see that at early times the agreement is poor as the neighboring oscillator which was initially at rest tries to synchronize with the perturbed center bond. This transient period lasts for about 20 units of time, after which, the LNE settles into its long-term quasi-periodic motion. The long-term motion is well described by Eq. (\ref{ep_ansatz}) using the coefficients found from Eqs. (\ref{a1_eq} \& \ref{b1_eq}) and captures a characteristic beating that we commonly observe in our simulations of LNEs. The agreement between the analytical and numerical results help assure us that the assumptions leading up to Eqs. (\ref{a1_eq} \& \ref{b1_eq}) are reasonable. Similar results were also reported in \cite{Kosevich2002} where the LNE is considered to be a composition of a stationary and moving breather. This interpretation is also consistent with the same beating phenomenon seen in Fig. \ref{r1_fit} as the moving breather oscillates back and forth within the LNE.

\section{LNE Emissions and Stability}
Since the LNE is not an exact breather solution as already discussed, some energy from the LNE will be exchanged with the rest of the chain. This energy will generally be much smaller than that contained within the LNE due to its inherent localization properties. The nature of the energy emissions from the LNE is determined by the strength of the excitation. Weakly excited LNEs will produce lower energy excitations in the neighboring oscillators such that the linear terms in Eq. (\ref{eom}) will dominate the dynamics. Hence these emissions are expected to be primarily acoustic in nature. Conversely, strongly excited LNEs will produce higher energy emissions such that the nonlinear terms in Eq. (\ref{eom}) will become more important. In these situations, we expect to see solitary waves and other nonlinear objects become more prevalent in the emission dynamics \cite{Kashyap2017}.
\par
Before proceeding, we should clarify what constitutes a ``weakly'' versus a ``strongly'' excited LNE. Our simulations show that even-parity LNEs initiated by a perturbation of $r_{0}(0)\lessapprox20$, which has a nonlinear component 400 times larger than the linear component in the original FPUT system, produce primarily acoustic emissions with frequencies confined to the phonon band for the majority of the LNE's lifetime. This is somewhat surprising, given the large amount of energy required for such a perturbation ($>45000$). Apparently, even fairly high energy LNEs which would possess internal dynamics approaching the purely nonlinear limit, still prefer to communicate with the rest of the chain through the phonon band and produce low energy, acoustic emissions up until delocalization. This is consistent with what was found in \cite{Mohan2011} where even a weak linear coupling (corresponding to both large $\beta$ and $A$ in this work) had a significant effect on the behavior and stability of an LNE. Hence, if one regards the presence of phonons effectively as ``noise,'' the observations bring to focus the possible importance of noise type effects on LNE stability \cite{Kenkre2002}. Considering these results, we will mainly focus on the ``weakly'' excited LNEs, which primarily produce acoustic emissions.
\par
Even though the LNE tends to communicate through the phonon band, this does not mean that nonlinear effects are unimportant, especially within the LNE itself. From Eq. (\ref{cubic_exp}), it is apparent that the terms $\cos([\omega_{a}-2\omega_{b}]t)$ and $\cos([\omega_{b}-2\omega_{a}]t)$ may introduce new frequencies within the phonon conduction band. In general, the localization property requires that both fundamental frequencies are above the phonon band, $\omega_{a}>\omega_{b}\ge2$, since excitations with frequencies less than 2 will strongly communicate with the phonon band \cite{Flach1994,Flach2005}; allowing energy to quickly flow from the LNE  into the normal modes of the system. Therefore, only the secondary frequency $|\omega_{a}-2\omega_{b}|$ is expected to lie within the phonon band. Meanwhile, the other secondary frequency $|\omega_{b}-2\omega_{a}|$ will be a high frequency greater than the two fundamental LNE frequencies $\omega_{a}$ \& $\omega_{b}$ and should be ignored according to the RWA.
\begin{figure}[t]
	\centering
	\includegraphics[width=\columnwidth]{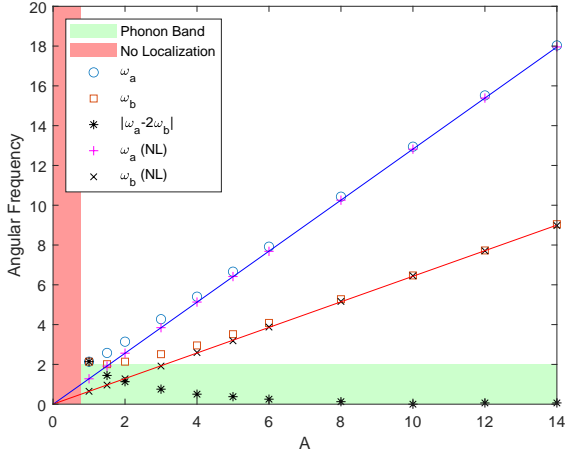}
	\caption{The frequencies $\omega_{a}$, $\omega_{b}$, and $|\omega_{a}-2\omega_{b}|$ of an even-parity LNE are plotted for varying initial amplitude $A$. For large $A$, the system approaches the purely nonlinear limit where the fundamental frequencies linearly scale with amplitude. The limiting behavior for $\omega_{a}$ and $\omega_{b}$ in the original system is illustrated by comparison with the frequencies from an LNE in the purely nonlinear case with the linear terms absent in Eq. (\ref{eom}), $\omega_{a}$ (NL) and $\omega_{b}$ (NL).}
	\label{fA}
\end{figure}
\par
In some cases, it is possible for the secondary frequency $|\omega_{a}-2\omega_{b}|$ to decouple from the phonon band. If $\omega_{a}\approx2\omega_{b}$ then the frequency $|\omega_{a}-2\omega_{b}|$ will move to the bottom of the phonon band. This occurs for even parity LNEs excited by a sufficiently large initial displacement $A\gtrapprox6$. This is shown in Fig. \ref{fA} where the frequencies $\omega_{a}$, $\omega_{b}$, and $|\omega_{a}-2\omega_{b}|$ are plotted for varying $A$. We see that as the amplitude is increased, the secondary frequency approaches zero. This corresponds with the scaling of the two fundamental frequencies in the strongly nonlinear limit. In this limit, time is scaled by the factor $1/A$; so frequency linearly scales with amplitude proportional to $A$. The frequency scaling in the nonlinear limit is found to follow $\omega_{a}=\left(1.283\pm0.001\right)A-\left(0.004\pm0.006\right)$ and $\omega_{b}=\left(0.6428\pm0.0032\right)A-\left(0.001\pm0.024\right)$ where the additive constant is close to zero as expected. It turns out that the ratio of the scaling between $\omega_{a}$ and $\omega_{b}$ in this limit is very close to 2 (specifically 1.996). This means that for large enough $A$, the even-parity LNE effectively decouples from the phonon band.
\par
The presence of the lower secondary frequency presents a pathway for energy to enter and leave the LNE through weak excitations in the phonon band. This could affect the stability of the LNE in two ways. If energy is allowed to leave the LNE, this may help the LNE move into a more stable configuration and increase its lifetime. Conversely, this energy pathway could make the LNE more sensitive to low energy background fluctuations, which would destabilize the LNE and decrease its lifetime.
\par
The stability of a breather type LNE is often addressed by studying the eigenvalues of the Floquet matrix corresponding to the linearized phase space flow around the LNE trajectory \cite{Flach2005,EnergyLocalization}. If we consider a perturbation to Eq. (\ref{eom}) of the form $r_{n}\rightarrow r_{n}+\epsilon_{n}$, then the dynamics of the perturbation to first order are described by the equations,
\begin{align}
\begin{split}
\dot{\epsilon}_{n}&=\delta_{n}\\
\dot{\delta}_{n}&=-\left(1+3r_{n-1}^{2}\right)\epsilon_{n-1}-\left(2+6r_{n}^{2}\right)\epsilon_{n}-\left(1+3r_{n+1}^{2}\right)\epsilon_{n+1}.
\end{split}
\label{pert}
\end{align}
Eqs. (\ref{pert}) can be written as the linear system,
\begin{align}
\frac{d}{dt}
\begin{pmatrix}
\vec{\epsilon}(t) \\
\vec{\delta}(t)
\end{pmatrix}
=
F(t)
\begin{pmatrix}
\vec{\epsilon}(t) \\
\vec{\delta}(t)
\end{pmatrix}.
\label{floqeq}
\end{align}
According to Floquet theory, if $F(t)$ is a $T$-periodic function then the solution to Eq. (\ref{floqeq}) is a linear combination of functions of the form $\phi_{n}(t)\exp(\sigma_{n} t)$ where the set of functions $\phi_{n}(t)$ are also $T$-periodic. These functions are the Floquet eigenfunctions of $F(t)$ while $\lambda_{n}=\exp(\sigma_{n} t)$ are the corresponding Floquet eigenvalues. The complex eigenvalues $\lambda_{n}$ of the Floquet matrix $F(t)$ determine the linear stability of the LNE. If all eigenvalues are of magnitude one, then the LNE is marginally stable. Otherwise, perturbations exists which will grow in time and the LNE is linearly unstable. The eigenvalues can be found by first numerically integrating Eqs. (\ref{pert}) over one period $T$. The resulting matrix $\left(\vec{\epsilon}(t_{i})\;\;\vec{\delta}(t_{i})\right)^{T}$ over the time series $\{t_{i}\}$ then has the set of eigenvalues corresponding to those of the Floquet matrix $\{\lambda_{n}\}$.
\par
\begin{figure}[t]
	\centering
	\includegraphics[width=\columnwidth]{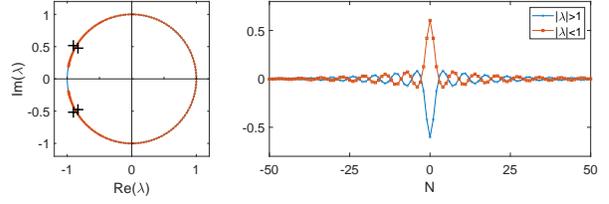}
	\caption{(Left) Floquet eigenvalues for an LNE with parameters $\omega_{a}=\omega_{b}\approx2.136$, $a_{1}=1$, $a_{2}=0.5$, and $b_{1}=b_{2}=0$. Two pairs of eigenvalues corresponding to instabilities in the LNE lie off of the unit circle and are indicated by ``+'' markers. Their approximate values are $-0.9044\pm0.5164i$ and $-0.8338\pm0.4761i$. (Right) Strain eigenvectors corresponding to the two pairs of eigenvalues with $|\lambda|\neq 1$.}
	\label{lf_floq}
\end{figure}
Unfortunately, this approach is generally limited to cases where it can be shown that the trajectory is closed and the motion periodic in time \cite{Zounes1998}; such that the LNE takes the form of a breather solution. This is only guaranteed if all pairs of frequencies comprising the LNE are commensurate ($f_{i}/f_{j}=N_{i}/N_{j}$ where both $N$ are integers) such that a common period exists ($T=N_{i}/f_{i}=N_{j}/f_{j}$) which defines the Floquet matrix in Eq. (\ref{floqeq}). Our numerical results indicate that the dominant frequencies $\omega_{a}$ \& $\omega_{b}$ are generally incommensurate for most values of $A$. One exception is near $A=1$ where the two frequencies begin to merge near the top of the phonon band (see Fig. \ref{fA}). In this case, only a single frequency is present $\omega_{a}=\omega_{b}\approx2.136$ and the corresponding parameters from Eq. (\ref{ep_ansatz}) are $a_{1}=1$, $a_{2}=0.5$, and $b_{1}=b_{2}=0$. The Floquet eigenvalues for this LNE are shown in Fig. \ref{lf_floq}. There are two pairs of eigenvalues which do not reside on the unit circle and correspond to instabilities in the LNE for the given parameters.
\par
To probe the question of stability for the multi-frequency LNEs with $A>1$, simulations are run for LNEs of varying amplitude in both a finite chain with fixed boundaries and an ``infinite" chain emulated by introducing damping at the ends of the chain. This prevents LNE emissions from returning at later times, similar to the case of an infinitely large system. The damping is applied to 10 oscillators towards each end of the chain and has the form $-\eta\dot{x}_{n}$ where $\eta$ starts at $0.1$ and doubles at each successive site. The particular form of the damping had little effect on the results as long as it was sufficient to absorb the bulk of the LNE emissions.
\par
The finite chain will allow energy to be reflected back from the boundaries and produce a weak background of excitations around the LNE. This background will be asymmetric so long as the LNE is not exactly centered in the chain. We expect that any interactions between the LNE and an asymmetric acoustic background will work to drive the system towards equilibrium. Hence, we expect the lifetime of the LNE to be shortened. Complementing this, the infinite chain will allow the LNE to release energy without this energy returning to the LNE. If the LNE releases energy in a way which stabilizes it, then the lifetime of the LNE should be longer than in the finite chain. We define the LNE lifetime by the time required for a given percentage of initial energy in the LNE to drop below a specified value. Since the energy cutoff chosen is somewhat arbitrary, we consider three possible values, $60\%$, $50\%$, and $40\%$, to illustrate how the measured lifetime may change with the cutoff value chosen.
\begin{figure}[t]
	\centering
	\includegraphics[width=\columnwidth]{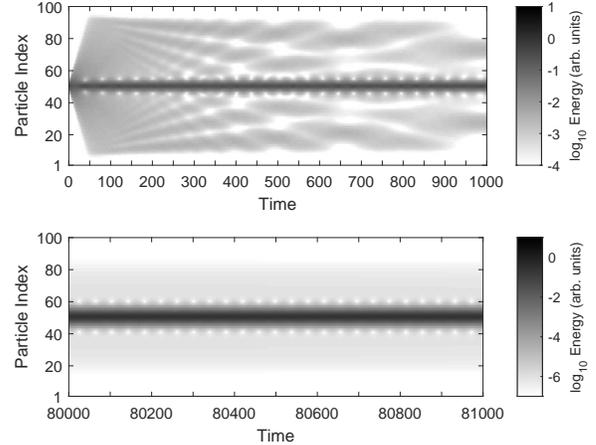}
	\caption{Energy per particle on log base 10 scale with damping near the boundaries. An even parity LNE is initiated with $r_{0}(0)=1.0$. Energy is quickly released by the LNE and propagates outward through the chain. At late times, emissions from the LNE become extremely weak.}
	\label{lne_damp}
\end{figure}
\par
For the case of an infinite chain, we find that the LNE can persist for extremely long times ($>10^{6}$). The LNE initially releases a train of excitations into the rest of the chain, which are absorbed by the damped oscillators near the boundaries. Over time, these emissions become quite weak and only a small amount of energy resides outside the LNE. This behavior is shown in Fig. \ref{lne_damp} where around $t=8\times10^{4}$, the emission energies are about seven orders of magnitude lower than the LNE energy. This implies that the stronger emissions at early times may be helping to stabilize the LNE; leading to its long lifetime. This behavior was observed for several amplitudes from $1.0$ to $10.0$. All had lifetimes exceeding what we could effectively simulate in a reasonable amount of time.
\begin{figure}[t]
	\centering
	\includegraphics[width=\columnwidth]{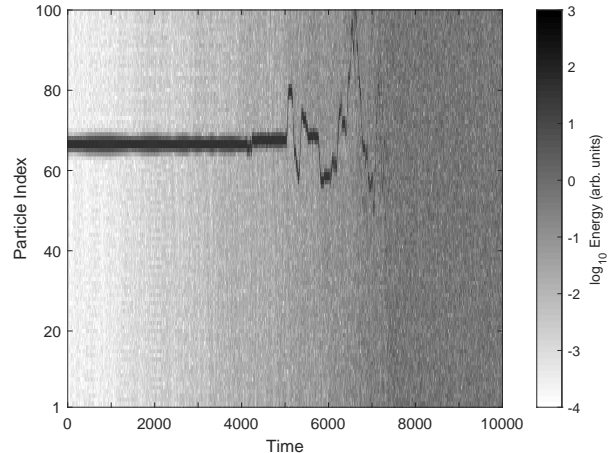}
	\caption{Energy per particle on log base 10 scale for a finite chain. An even parity LNE is initiated with $r_{0}(0)=3.0$. Energy is quickly released by the LNE and travels back and forth through the chain. This eventually leads to the LNE destabilizing and dissipating.}
	\label{lne_des}
\end{figure}
\par
\begin{figure}[t]
	\centering
	\includegraphics[width=\columnwidth]{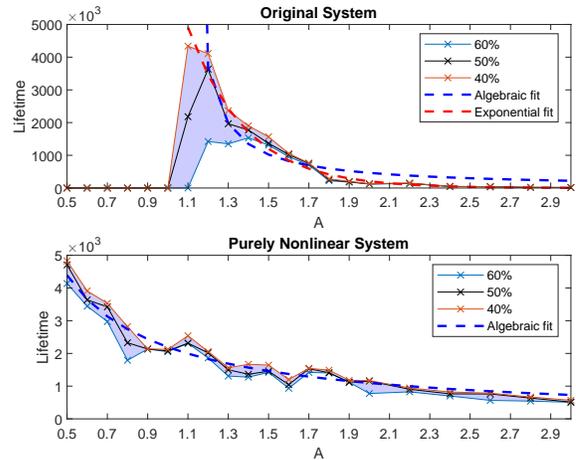}
	\caption{The lifetime of the LNE is plotted against the amplitude $A$ for a chain of 100 particles. The top panel considers the original system defined by Eq. (\ref{eom}) while the bottom panel considers a purely nonlinear system with the linear terms in Eq. (\ref{eom}) absent. The lifetime is measured by the time required for a given percentage of initial energy in the LNE to drop below a specified value, either $60\%$, $50\%$, or $40\%$.}
	\label{lne_lt}
\end{figure}

For the case of the finite chain, the lifetime of the LNEs is generally much shorter. Which is a consequence of energy reflected by the boundaries colliding with and destabilizing the LNE. Such a destabilization event is shown in Fig. \ref{lne_des}. The extent to which this occurs is dependent on both the amplitude of the LNE and the size of the chain, with longer chains allowing for longer lived LNEs. Fig. \ref{lne_lt} shows the lifetime of a LNE as a function of amplitude for a 100 particle chain with the LNE centered across particles 66 \& 67. The top panel shows the lifetime for the original system defined by Eq. (\ref{eom}) while the bottom panel considers a purely nonlinear system with the linear terms in Eq. (\ref{eom}) absent.
\par	
We see that there is a sudden increase in the lifetime of the LNE for amplitudes greater than one in the original system. This corresponds with the point where the secondary frequency enters the phonon band in Fig. \ref{fA}. Note that the LNE lifetime in the original system is roughly $2000$ times larger than in the corresponding purely nonlinear system for amplitudes around $1.2$. Since the purely nonlinear system lacks a phonon band, it would appear that the dramatic increase in lifetime is associated with the presence of secondary frequency within the band. We therefore conclude that communication between the LNE and the phonon band through the secondary frequency is helping to stabilize the LNE and that a possible increase in the LNE's susceptibility to noise from the phonon band is less important \cite{Mohan2011}.
\par 
After the initial increase in lifetime for $A>1$, larger values of $A$ generally shorten the lifetime due to the larger energy effectively speeding up the dynamics. This is illustrated by the lifetime dependence in the purely nonlinear system which should scale with the inverse of amplitude. A fit to this algebraic inverse scaling with $A$ is shown for both systems in Fig \ref{lne_lt}. The fit appears good for the purely nonlinear system, giving a lifetime $T$ that scales as $T\approx\left(2200\pm100\right)/A$. For the original system however, the inverse scaling overestimates the lifetime at larger amplitudes. A better fit is given by an exponential scaling law $T\approx\left(2.4\pm0.8\right)\times10^{8}\cdot\exp\left[-\left(3.5\pm0.6\right)A\right]$. However, the origin of this exponential scaling is not immediately apparent. We conclude this section by noting that in terms of the original system variables where $x_{1}-x_{0}$ has a dimensional amplitude $\mathcal{A}$, we find that the LNE lifetime is dependent on both the amplitude and nonlinearity parameter such that the lifetime is maximized around $\mathcal{A}\sqrt{\beta}\approx1.1$.

\section{Extension to higher dimensional lattices}
\begin{figure}[t]
	\centering
	\includegraphics[width=\columnwidth]{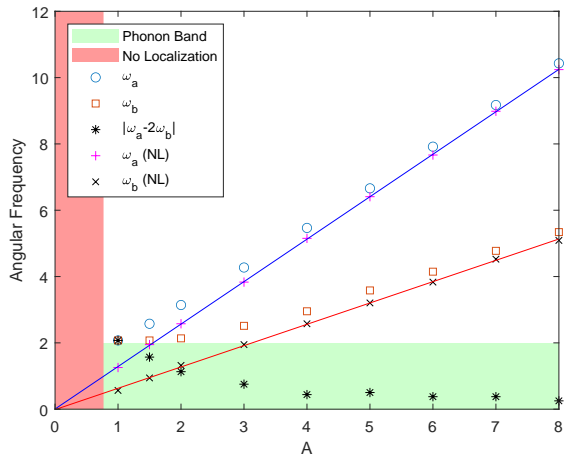}
	\caption{The frequencies $\omega_{a}$, $\omega_{b}$, and $|\omega_{a}-2\omega_{b}|$ of an even-parity LNE are plotted for varying initial amplitude $A$ in a cubic lattice. The limiting behavior for $\omega_{a}$ and $\omega_{b}$ in the original system is illustrated by comparison with the frequencies from an LNE in the purely nonlinear case with the linear terms absent in Eq. (\ref{eom}), $\omega_{a}$ (NL) and $\omega_{b}$ (NL).}
	\label{3d_fA}
\end{figure}

Similar trends in the LNE lifetime are also observed in higher dimensions. Using the discrete element modeling software \textsc{lammps} \cite{Plimpton1995}, LNEs were initiated in a cubic FPUT lattice 50 particles on a side. Localization is again only seen for sufficiently large initial perturbations \cite{Jackson1990}. The frequencies of interest are shown for the cubic lattice case in Fig. \ref{3d_fA}. Similar to the 1D chain, the frequency scaling for the cubic lattice in the nonlinear limit is linear with amplitude and follows $\omega_{a}=\left(1.281\pm0.007\right)A+\left(0.005\pm0.008\right)$ and $\omega_{b}=\left(0.6432\pm0.0132\right)A-\left(0.013\pm0.037\right)$. Our numerical results show that the amplitude cutoff for localization is very similar in the cubic lattice to that in the original 1D chain. Furthermore, the lower frequency $\omega_{b}$ is again found to stay above 2, indicating that the same constraints on localization in the 1D chain must apply to higher dimensional lattices.
\par
One explanation for these similarities is that the energy released by the LNE tends to move along a chain of oscillators in the cubic lattice coincident with the LNE. This would lead to similar dynamics in both the 1D and 3D models. Our numerical results support this idea. We find that at late times but before the LNE delocalizes, about $3\%$ of the initial energy has dissipated from the LNE and into the adjacent oscillators along the same axis as the LNE. Meanwhile, about $0.5\%$ of the energy has dissipated into the rest of the entire lattice. This observation is consistent with existing literature for 2D lattices which show that localized excitations move and disperse energy in a preferred direction along symmetries of the lattice \cite{Yi2009,Marin2000,Marin1998}.

\begin{figure}[t]
	\centering
	\includegraphics[width=\columnwidth]{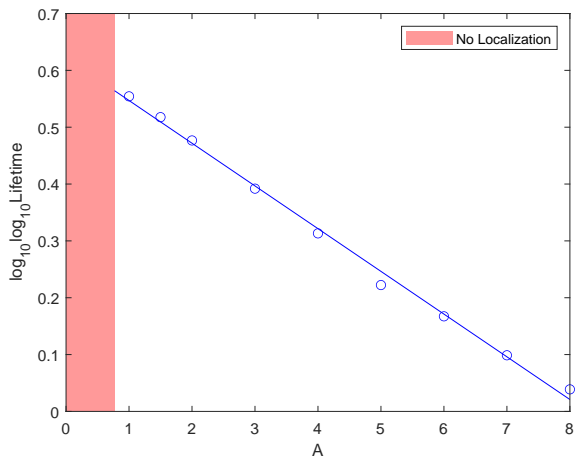}
	\caption{The lifetime of the LNE in a cubic FPUT lattice on a log-log base 10 scale is plotted against the LNE amplitude.}
	\label{3d_lne_lt}
\end{figure}

\par
LNEs in the cubic lattice do not show the dramatic increase in lifetime that was seen in the 1D chain at the lowest amplitudes, rather, the lifetime monotonically decreases with amplitude. The extra degrees of freedom afforded in the cubic lattice likely offer more pathways for energy to be exchanged between the LNE and the rest of the system. This would allow the lower amplitude LNEs to readily stabilize at early times, similar to how a secondary frequency in the phonon band helped to stabilize the LNEs in the 1D chain. We expect that an increased number of degrees of freedom should lead to a strong decrease in the LNE lifetime for increasing amplitude. Large energy LNEs are able to more effectively excite off axis oscillations near the LNE which could quickly shunt energy away from the LNE. Numerical results show that the LNE lifetime in the 3D lattice does indeed quickly decrease for large amplitudes. We find that the LNE lifetime $T$ is well described by the double log dependence $\log_{10}\log_{10}(T)\approx -(0.075\pm0.005)A+(0.62\pm0.02)$ as illustrated in Fig. \ref{3d_lne_lt} where an energy cutoff of $50\%$ is used. In terms of the variables in Eq. (\ref{fput}), this dependence is $\log_{10}\log_{10}(T)\approx -(0.075\pm0.005)\mathcal{A}\sqrt{\beta}+(0.62\pm0.02)$: indicating a rapid decrease in the lifetime with increasing amplitude and/or increasing nonlinear strength.

\section{Conclusion}
The dynamics of LNEs in the $\beta$-FPUT system are discussed in terms of the dependence of the LNE lifetime on amplitude, and by extension the strength of the nonlinearity. We first show how the interaction of the two fundamental frequencies within the LNE leads to the production of secondary frequencies which can lie within the phonon band. These frequencies are seen for LNEs in both the 1D chain and 3D cubic lattice. The presence of the secondary frequency within the phonon band is related to the problem of LNE stabilization since the LNE emissions studied here are of low energy and hence primarily acoustic. The question of how a resonance in the phonon band may interact with the LNE and the acoustic background is explored by comparing the lifetime of the LNE in an emulated infinitely long chain, a finite chain, and a purely nonlinear chain. We observe that LNEs in the infinite chain are extremely long lived while LNEs in the finite chain have lifetimes dependent on the LNE amplitude $\mathcal{A}$ and the nonlinear strength $\beta$, with the longest lived LNEs satisfying $\mathcal{A}\sqrt{\beta}\approx1.1$.
\par
The LNE lifetime dramatically increases once the lower secondary frequency enters the phonon band, leading us to conclude that the location of this frequency within the band opens a pathway for low energy excitations to enter and leave the LNE at early times and improve its stability. LNEs without this energy pathway, either in the purely nonlinear system or with $\mathcal{A}\sqrt{\beta}\leq1$, are less stable and prone to delocalize when energy reflected back from the boundaries interacts with the LNE. We find that after the initial increase in LNE lifetime, the lifetime decreases exponentially with amplitude rather than by an inverse relationship as in the purely nonlinear system. In the cubic lattice, we find that LNEs have lifetimes that scale with amplitude via a double log relationship. The precise origin of these scaling relationships are still unknown, which we believe presents an interesting topic for future work.
\section*{Acknowledgement}
We would like to thank Dr. Sergej Flach for his helpful comments and advice on clarifying some of the arguments presented in this work.

\bibliographystyle{ieeetr}
\bibliography{ref}
\end{document}